\documentclass[a4paper]{article}

\usepackage{ISCSLP2022}
\usepackage{multirow}
\usepackage{algorithmic}
\usepackage{url}
\usepackage{algorithm,color,cite}

\title{LCM-SVC: Latent Diffusion Model Based Singing Voice Conversion with Inference Acceleration via Latent Consistency Distillation }
\name{Shihao Chen$^{1,2}$, Yu Gu$^{2}$, Jianwei Cui$^{1,2}$, Jie Zhang$^{1}$\thanks{This work was done at Tencent AI Lab as an internship by Shihao Chen. (Correspondence: jzhang6@ustc.edu.cn)},Rilin Chen$^2$, Lirong Dai$^{1}$}
\address{
  $^{1}$NERC-SLIP, University of Science and Technology of China (USTC), China \\
  $^{2}$ Tencent AI Lab
}
\email{shchen16@mail.ustc.edu.cn, colinygu@tencent.com, jzhang6@ustc.edu.cn}

\begin{document}
\maketitle
\begin{abstract}
Any-to-any singing voice conversion (SVC) aims to transfer a target singer's timbre to other songs using a short voice sample. However many diffusion model based any-to-any SVC methods, which have achieved impressive results, usually suffered from low efficiency caused by a mass of inference steps.
 In this paper, we propose LCM-SVC, a latent consistency distillation (LCD) based latent  diffusion model (LDM) to  accelerate inference speed. 
 We  achieved one-step or few-step inference while maintaining the high performance by distilling a pre-trained LDM based SVC model, which had the advantages of  timbre decoupling and sound quality. Experimental results show that our proposed method can significantly reduce the inference time and largely preserve the sound quality and timbre similarity comparing with other state-of-the-art SVC models. Audio samples are available at \url{https://sounddemos.github.io/lcm-svc/}.
\end{abstract}

\noindent\textbf{Index Terms}: Singing voice conversion, latent diffusion model, consistency distillation

\section{Introduction}
Singing voice conversion (SVC) is an emerging audio editing application that aims to transfer the timbre of a target singer to another piece of singing, such as allowing a celebrity to sing a song we have composed ourselves. Unlike singing voice synthesis
 \cite{bytesing,cui2024sifisinger}, SVC does not require the input of musical scores or lyrics and  it can be accomplished with just singing voice input. Due to the scarcity of paired data, current SVC models primarily focus on  the task of decoupling information in singing, e.g., content, timbre, pitch. By training a reconstruction model provided relevant clues, these features can be reassembled into singing and some can be replaced to achieve timbre conversion or pitch correction.  There were many works that have achieved good results on SVC task, such as models based on generative adversarial networks \cite{polyak2020unsupervised,liu2021fastsvc,zhou2022hifi}, models based on Variational AutoEncoder (VAE) \cite{luo2020singing}, models based on Diffusion \cite{liu2021diffsvc,lu2024comosvc} and end-to-end model So-VITS-SVC\footnote{\url{https://github.com/PlayVoice/so-vits-svc-5.0/tree/bigvgan-mix-v2}}. Among these systems, diffusion models exhibit a superiority in sound quality and timbre similarity, particularly using latent diffusion model (LDM) \cite{chen2024ldmsvc}.

The major drawback of the diffusion model based methods is the long required inference time. The inference iterations of diffusion models can reach up to 100 or even more than 1000, which is impractical for real-world applications. Some efforts were thus made to accelerate the inference, such as Denoising Diffusion Implicit Models (DDIM) \cite{song2020denoising}, Diffusion Probabilistic Models Solver (DPM-Solver) \cite{lu2022dpm}, etc.
Recently, the emergence of consistency model (CM) provides a new proposal for accelerating the inference of diffusion, where the goal is to ensure that the output at each step of the diffusion's denoising process remains consistent \cite{song2023consistency}. By learning consistency mappings that maintain point consistency on Order Linear Equations (ODE)-trajectory, CM achieves one-step generation and thus avoids computationally intensive iterations.
Further, the advent of latent consistency model (LCM) \cite{luo2023latent} suggests that applying consistency distillation in the latent space can yield superior results, which can also improve the efficiency of diffusion  models, meanwhile maintaining high-quality outputs.

In this paper, we propose an SVC method (abbreviated by LCM-SVC) using latent consistency distillation (LCD) strategy on the basis of an LDM SVC model. This method efficiently converts a pre-trained LDM into an LCM by solving an augmented Probability Flow ODE (PF-ODE). Initially, we train a So-VITS-SVC model as the VAE structure to extract hidden latent variables and then a teacher model based on LDM following  a classifier guidance scheme. We then utilize the LCD method to distill the model, facilitating few-step or even one-step inference while preserving audio quality similarly to the teacher model. We further apply a skipping-step technique to accelerate the convergence of model training. Experimental results indicate that LCM-SVC can only incur a slight loss in audio quality using one-step inference iteration. More importantly, in case of using 2-step or 4-step inference, the obtained audio quality is comparable to that of the teacher model and the inference time is significantly reduced, which satisfies the efficiency request of practical applications. 
The rest of this paper is organized as follows. Section 2 outlines the proposed LCM-SVC method. Experiments are presented in Section 3. Finally, Section 4 concludes this work.
\section{Method}
 \begin{figure*}[t]
  \centering
  \includegraphics[width=\textwidth]{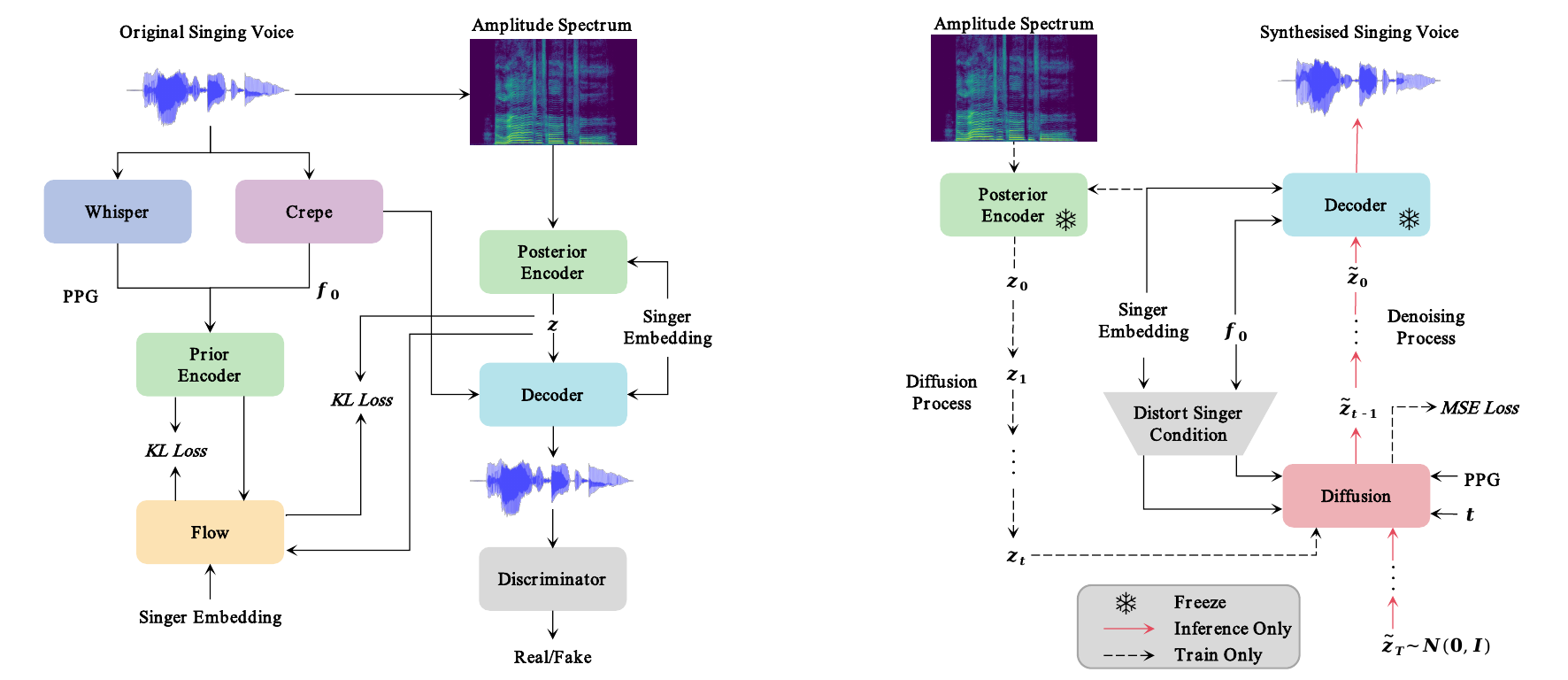}
  \caption{Left: Pre-training procedure of So-VITS-SVC; Right: Training procedure of LCM-SVC Teacher.}
  \label{ldm-svc}
\end{figure*}
\subsection{Latent Diffusion Model for SVC}
The VAE model is firstly pre-trained to extract hidden latent vectors using So-VITS-SVC, which is a variant of VITS \cite{kim2021conditional}, consisting of three key components: posterior encoder, prior encoder and decoder as depicted in Figure 1. 
The posterior encoder composed of non-causal WaveNet \cite{oord2016wavenet} residual blocks models the posterior distribution $p(z|y,e)$ of the hidden representation $z$ from the linear spectrograms  $y$ generated from the  singing waveforms,  where singer embedding $e$ is extracted by an additional speaker verification model.
The prior encoder, constructed with a multi-layer Transformer \cite{vaswani2017attention}, estimates the prior distribution $p(z|x,f_0,e)$, where $x$ and $f_0$ represent the phonetic posteriorgrams (PPG) and fundamental frequency (F0) respectively. A normalizing flow \cite{papamakarios2021normalizing} is utilized to transform the distributions of the prior and posterior encoders.
The BigVGAN-based \cite{lee2022bigvgan} decoder generates the singing waveform from the latent representation $z$, using a neural source filter (NSF) \cite{wang2019neural} scheme with F0s to enhance pitch accuracy. After training, the posterior encoder and decoder are used to extract hidden latent variables for LDM and synthesize waveforms from the hidden latent vectors predicted by LDM.

The LDM is adopted as the probabilistic model that fit the hidden distribution $p(z_{0})$ by denoising in data latent space from the pre-trained VAE model.  Denoising diffusion probabilistic model (DDPM) \cite{ho2020denoising} is used, which consists of forward and denoising processes. During the LDM training, the singer's timbre $e$ and the linear spectrogram of the singing voice $y$ are used as inputs to the posterior encoder $\mathcal{E}(\cdot)$, yielding the latent variable $z_0=\mathcal{E}(y,e)$. 
In the forward process, the original data distribution is transformed into a standard Gaussian distribution by gradually adding noise according to a fixed schedule $\beta_1,\dots,\beta_T$, where $T$ represents the total time steps. This process includes a transition from $z_{0}$ to $z_t$ following a Markov chain, where the conditional distribution $q(z_t|z_{t-1})$ is defined as a Gaussian distribution, i.e., $q(z_t|z_{t-1}) = \mathcal{N}(z_t;\sqrt{1-\beta_t}z_{t-1}, {\beta_t}\mathbf{I})$.

The reverse process, denoted by $\theta$, functions as a denoising mechanism to suppress noise and recover the original data. The denoising distribution, $p_{\theta}(z_{t-1}|z_t)$, is a conditional Gaussian distribution. As such, we iteratively sample target data $z_0$ from Gaussian noise for $t = T, T-1, \ldots, 1$, with $z_{t-1}$ sampled based on $p_{\theta}(z_{t-1}|z_t)$. The LDM uses $z_t$ and $t$ as inputs, together with conditional inputs like the singer's timbre $e$, fundamental frequency $f_0$, and PPG $x$. We also use a singer guidance method, specifically Speaker Condition Layer Normalization (SCLN) \cite{wu2021cross} for PPG feature normalization, and a classifier-free guidance method \cite{ho2022classifier} for model training to better decouple timbre information. The step-wise output $z_{t-1}$ of denoising  is calculated as:
\begin{equation}z_{t-1} = \frac{1}{\sqrt{\alpha_t}} \left( z_t - \frac{1-\alpha_t}{\sqrt{1-\bar{\alpha}_t}} \epsilon_{\theta}(z_t, t, x, f_0, e) \right) + \sigma_t \epsilon,
\end{equation}
where $\epsilon \sim \mathcal{N}(0, I)$, $\alpha_t = 1 - \beta_t$ and $\bar{\alpha}_t = \prod_{s=1}^t \alpha_s$. For the detailed derivation of (1), please refer to \cite{ho2020denoising}. The configuration of the considered diffusion model keeps consistent with that of DiffSVC. The training loss of the LDM $\epsilon_\theta$ is defined as the mean squared error (MSE) in the noise space: 
\begin{equation}
    \mathcal{L}_{\rm LDM}=||\epsilon-\epsilon_\theta(z_t, t, x, f_0, e)||_2^2.
\end{equation}

During inference, we input source PPG $x_{src}$, replace the singer timbre with $e_{tar}$ and adjust $f_0$ with the target attribute. We sample random Gaussian white noise to obtain $\widetilde{z}_0$ using the denoising process. Then, we input $\widetilde{z}_0$, $e_{tar}$, and $f_0$ into the pre-trained decoder $\mathcal{D}(\cdot)$ to generate the audio waveform.
\subsection{Latent Consistency Distillation}
 \begin{figure}[t]
  \centering
  \includegraphics[width=0.5\textwidth]{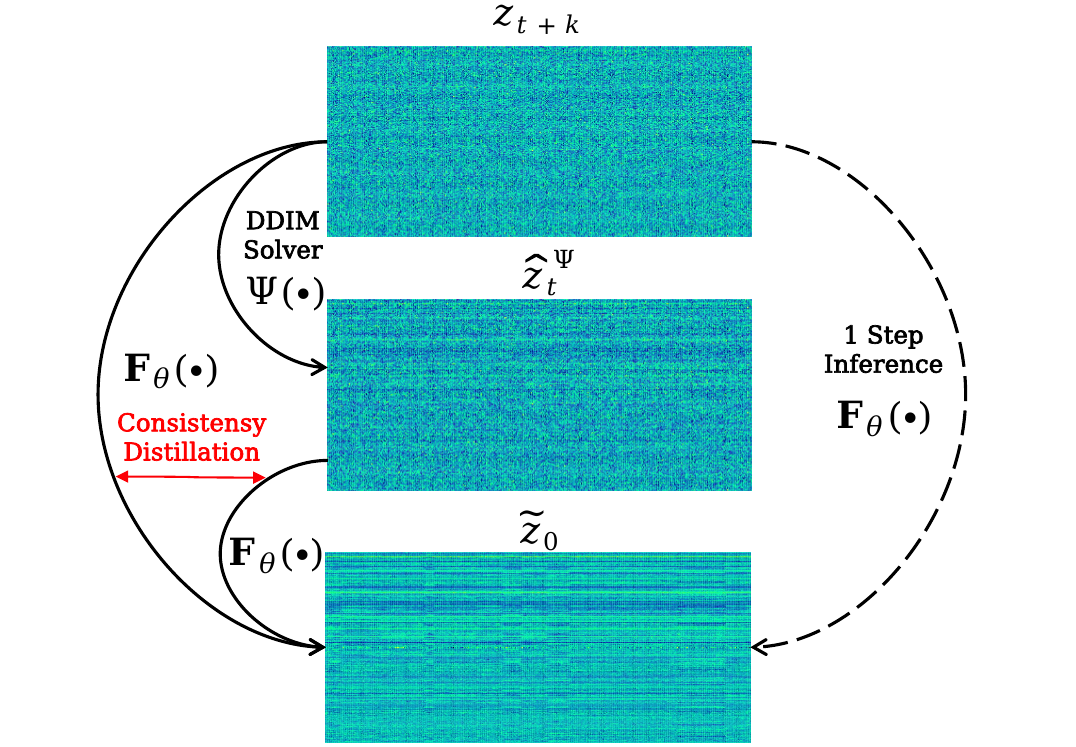}
  \caption{The Training and Inference procedure of LCD.}
  \label{cm}
\end{figure}
Following \cite{luo2023latent}, we use latent consistency distillation (LCD) method to accelerate the reverse process of LDM. The Consistency Model (CM)~\cite{song2023consistency} facilitates one-step or few-step generation. The fundamental concept of the CM is to learn a function capable of mapping any points on a trajectory of the Probability Flow ODE (PF-ODE) back to the trajectory's origin, which is essentially the solution of the PF-ODE. The general process of LCD is shown in Figure \ref{cm}.

As the purpose of LCD is to define a function such that the target is the same for each time step $t$, in order to maintain the accuracy of the prediction, the function $\textbf{F}_{\theta}$ is defined as:
\begin{equation}
    \resizebox{\linewidth}{!}{$\textbf{F}_{\theta}(z_t, x,f_0,e, t) = c_{\text{skip}}(t)z_t + c_{\text{out}}(t)\left(\frac{z_t - \sigma_{t}\epsilon_{\theta}(z_t, x,f_0,e, t)}{\alpha_t}\right)$}, \notag
\end{equation}
where $c_{\text{skip}}(t)$ and $c_{\text{out}}(t)$ are differentiable functions with $c_{\text{skip}}(0)=1$ and $c_{\text{out}}(0)=0$. In order to facilitate consistency distillation, we apply a skipping-step method to expedite the convergence process. Specifically, we use an ODE solver, denoted as $\Psi(\cdot)$, to predict $\hat{z}_{t}^{\Psi}$, which is based on $z_{t+k}$ with $k$ being the interval.
\begin{equation}
    \resizebox{\linewidth}{!}{$\hat{z}_{t}^{\Psi}= (1+\omega)\Psi(z_{t+k},t+k,t,x,f_0,e)-\omega\Psi(z_{t+k},t+k,t,x,\varnothing)$}, \notag
\end{equation}
where $\omega$ is the guidance weight and $\Psi$ the DDIM solver. To enforce the self-consistency property, a target model $\theta^-$ is maintained. This model is updated with the exponential moving average (EMA) of the parameter $\theta$ that we aim to learn. This can be mathematically represented as:
\begin{equation}
\theta^{-} \leftarrow\mu\theta^{-}+(1-\mu)\theta.
\end{equation}
To maintain the consistency of the model's output, we  define a loss function to constrain the consistency of the outputs at different steps, given by:
\begin{equation}
 \resizebox{\linewidth}{!}{$\mathcal{L}(\theta,\theta^{-};\Psi)= d(\mathbf{F}_\theta(z_{t+k},x,f_0,e,t+k),\mathbf{F}_{\theta^-}(\hat{z}_{t}^{\Psi},x,f_0,e,t))$}, \notag
\end{equation}
where $d(\cdot,\cdot)$ measures the distance, e.g., the often-used l2 norm. During back propagation, only $\theta$ is updated, and note that $\theta^- $is updated through EMA. The ODE solver $\Psi$ uses the frozen parameters obtained by the pre-trained LDM. The LCD training procedure is summarized in Algorithm 1. After the completion of training, the sample quality can be significantly improved via a sequence of denoising and noise injection steps. In the $n$-th iteration, we initiate a noise-injecting forward process on the previously predicted sample $\widetilde{z}_0$ as $\widetilde{z}_{\tau_{n}}\sim\mathcal{N}(\alpha_{\tau_{n}}\widetilde{z}_0,\sigma_{\tau_{n}}I)$ with
$\tau_{n}$ denoting a decreasing sequence of time steps. The key steps of the inference procedure are outlined in Algorithm 2. For more detailed information, please refer to \cite{luo2023latent}.


\begin{algorithm}[t]
\caption{Consistency Distillation of LCM-SVC.}
\begin{algorithmic}[1]
\label{alg:training}
\REQUIRE Initial model parameter $\theta$; training set $D_{train} = \{(x,{f_0},e,y)\}_{m=1}^M$; pretrained So-VITS-SVC posterior encoder $\mathcal{E(\cdot)}$; EMA rate $\mu$; noise schedule $\alpha_{t},\sigma_{t}$; guidance weight $\omega$; ODE solver $\Psi(\cdot)$; skipping interval $k$; distance metric $d(\cdot,\cdot)$; learning rate $\eta$.
\STATE $\theta^{-}\leftarrow\theta$
\REPEAT
\STATE Sample $(x, f_0, e, y)$ from $D_{train}$;
\STATE $z_0\leftarrow\mathcal{E}(y,e)$;
\STATE Sample $t\sim$ Uniform$(\{1, \cdots, T-k\})$;
\STATE Sample $z_{t+k}\sim\mathcal{N}(\alpha_{t+k}z_0;\sigma_{t+k}I)$;
\STATE $\hat{z}_{t}^{\Psi}\leftarrow  (1+\omega)\Psi(z_{t+k},t+k,t,x,f_0,e)$\\\qquad\qquad$-\omega\Psi(z_{t+k},t+k,t,x,\varnothing)$
\STATE $\mathcal{L}(\theta,\theta^{-};\Psi)\leftarrow d(\mathbf{F}_\theta(z_{t+k},x,f_0,e,t+k),$\\\qquad\qquad\qquad~~~~~~$\mathbf{F}_{\theta^-}(\hat{z}_{t}^{\Psi},x,f_0,e,t))$;
\STATE $\theta\leftarrow \theta - \eta\nabla_\theta(\theta,\theta^{-})$;
\STATE $\theta^{-} \leftarrow$ stopgrad$(\mu\theta^{-}+(1-\mu)\theta)$;
\UNTIL convergence.
\end{algorithmic}
\end{algorithm}

\begin{algorithm}[h]
\caption{Multi-step Inference of LCM-SVC.}
\begin{algorithmic}[1]
\label{alg:infer}
\REQUIRE Latent Consistency Model $\mathbf{F}_\theta(\cdot)$; source singer 
PPG $x_{src}$; target singer embedding $e_{tar}$; modified $f_0$; pretrained So-VITS-SVC decoder $\mathcal{D}(\cdot)$; noise schedule $\alpha_{t},\sigma_{t}$; sequence of timesteps $\tau_{1}>\tau_{2}>...>\tau_{N-1}$.\STATE Sample $\widetilde{z}_T\sim\mathcal{N}(0,I)$;
\STATE $\widetilde{z}_0\leftarrow \mathbf{F}_\theta(\widetilde{z}_T,x_{src},f_0,e_{tar},T)$;
\FOR{$n=1,2,...,N-1$}
\STATE $\widetilde{z}_{\tau_{n}}\sim\mathcal{N}(\alpha_{\tau_{n}}\widetilde{z}_0,\sigma_{\tau_{n}}\mathbf{I})$;
\STATE $\widetilde{z}_0\leftarrow \mathbf{F}_\theta(\widetilde{z}_{\tau_{n}},x_{src},f_0,e_{tar},\tau_{n})$;
\ENDFOR
\RETURN $\mathcal{D}(\widetilde{z}_0,f_0,e_{tar})$;
\end{algorithmic}
\end{algorithm}

\section{Experiment}

\subsection{Experimental Setup}

\begin{table*}[ht]
\caption{Subjective indicators (SMOS, NMOS) and objective indicators (SSIM, FPC) of comparison methods.}
\renewcommand{\arraystretch}{1}
\setlength\tabcolsep{2pt} 
\centering
\resizebox{\textwidth}{!}{%
\begin{tabular}{@{}l|cccccccc@{}}
\toprule
\multirow{2.5}{*}{Method} & \multicolumn{4}{c|}{Seen Singer}                          & \multicolumn{4}{c}{Unseen Singer}      \\ \cmidrule(l){2-9} 
 &
  \multicolumn{1}{c}{NMOS} &
  \multicolumn{1}{c}{SMOS} &
  \multicolumn{1}{c}{SSIM} &
  \multicolumn{1}{c|}{FPC} &
  \multicolumn{1}{c}{NMOS} &
  \multicolumn{1}{c}{SMOS} &
  \multicolumn{1}{c}{SSIM} &
  \multicolumn{1}{c}{FPC} \\ \midrule
DiffSVC      & $3.821\pm0.075$  & $3.950\pm0.091$& $0.593$ & \multicolumn{1}{c|}{$0.944$}  &$3.689\pm0.085$ & $3.906\pm0.101$ & $0.586$ & $0.942$ \\
So-VITS-SVC   & $3.872\pm0.070$  & $3.892\pm0.099$& $0.639$ & \multicolumn{1}{c|}{$\bf{0.946}$} &$3.753\pm0.087$ & $3.753\pm0.087$ & $0.600$ & $0.942$ \\
CoMoSVC   & $3.855\pm0.068$  & $3.953\pm0.096$& $0.639$ & \multicolumn{1}{c|}{$0.945$} &$3.700\pm0.086$ & $3.922\pm0.095$ & $0.600$ & $\bf{0.943}$ \\
 LCM-SVC-T &$\bf{3.975\pm0.066}$  & 
$\bf{4.139\pm0.087}$ & $\bf{0.702}$ & \multicolumn{1}{c|}{$0.945$}  &$\bf{3.794\pm0.086}$ & $\bf{4.061\pm0.105}$ & $\bf{0.663}$ & $0.942$\\
 LCM-SVC   &  $3.827\pm0.078$ & $4.125\pm0.090$  & $0.690$ &  \multicolumn{1}{c|}{$~~0.943~~$}& $3.689\pm0.088$ & $3.922\pm0.103$ & $0.652$& $~~0.941~~$ 
\\
 \bottomrule
\end{tabular}
}
\label{result_eer}
\end{table*}
\begin{table}[h]
\caption{Subjective evaluation results of LCM-SVC-$t$ inference using $t$ iterations.}
\setlength\tabcolsep{2pt} 
\centering
\resizebox{0.4\textwidth}{!}{%
\begin{tabular}{@{}l|c|c@{}}
\toprule
\multicolumn{1}{l}{Method} &
  \multicolumn{1}{|c}{NMOS} &
  \multicolumn{1}{|c}{SMOS}  \\ \midrule
\multicolumn{3}{c}{Seen Singer}\\ \midrule
 LCM-SVC-4~~ &$~~\bf{3.880\pm0.073}~~$  & $~~4.119\pm0.090~~$ \\
 LCM-SVC-2 &$3.877\pm0.075$  & $\bf{4.136\pm0.087}$ \\
 LCM-SVC-1   &  $3.827\pm0.078$ & $4.125\pm0.090$ \\ \midrule
\multicolumn{3}{c}{Unseen Singer}\\ \midrule
 LCM-SVC-4 &$\bf{3.767\pm0.081}$  & $4.031\pm0.103$ \\
 LCM-SVC-2 &$3.711\pm0.087$  & $\bf{4.036\pm0.102}$ \\
 LCM-SVC-1   &  $3.689\pm0.088$ & $3.922\pm0.103$
\\
 \bottomrule
\end{tabular}
}
\label{lcm}
\end{table}

\vspace{-0.2cm}
For evaluation, we utilize the OpenSinger dataset~\cite{huang2021multi}, with a comprehensive collection of Chinese singers. The dataset encompasses 74 singers, including 27 males and 47 females, and comprises a total of 52 hours of recordings.  To evaluate the conversion performances of different systems, two scenarios depending on if the target singers are included in the training dataset, i.e., seen and unseen singers are conducted. In the test set, 4 male and 4 female singers are randomly selected as unseen singers. For the seen singers' case, 8 sentences from each of the remaining 66 singers are chosen for testing, while the rest of the songs are used for training.

For comparative analysis, we trained So-VITS-SVC, DiffSVC, CoMoSVC and LCM-SVC teacher model (LCM-SVC-T). To ensure a fair comparison, we extract the PPG for each model using Whisper~\cite{radford2023robust}, and the F0 using CREPE~\cite{kim2018crepe}. Singer embeddings are obtained from the pre-trained model of CAM++~\cite{wang2023cam++}, an open-source project for speaker verification\footnote{\url{https://modelscope.cn/models/iic/speech_campplus_sv_zh-cn_16k-common/summary}}. In experiments, all audio files are resampled to 32kHz. Each model is trained for 400k steps  with a total batch size of 32. For the F0 embedding in the diffusion condition, we first quantize the Log-F0 features into 256 bins, followed by a pass through a melody embedding lookup table. During inference, we adjust the F0 of the original singer by calculating the mean F0 values of both the target and source singers within the voice segment, denoted as $\mathrm{mean}({f_0}^{src}_v)$ and $\mathrm{mean}({f_0}^{tar}_v)$, respectively. We then scale ${f_0}^{src}$ by the ratio of these mean values to obtain the modified ${f_0}$ as ${f_0}^{src} \times \frac{\mathrm{mean}({f_0}^{tar}_v)}{\mathrm{mean}({f_0}^{src}_v)}$. For the singer guidance method, we set the distortion probability $p_{\text{uncond}}$ to 0.1 during training, and the guidance weight $w$ to 0.3 during inference, in accordance with~\cite{ho2022classifier}. The total number of diffusion steps is set to 100 (i.e., $T = 100$). The noise schedule $\beta$ is linearly distributed, starting from a minimal value of 1e-4 and reaching up to $0.06$, following the configuration used in the DiffSVC model. For CoMoSVC, we use the same configuration as \cite{lu2024comosvc}, that is, the number of steps for training the teacher model is 50. As for the consistency distillation part, both LCM-SVC and CoMoSVC adopt the same hyper-parameters, where $\mu$ = 0.95 and the learning rate is set to 5e-5. The interval $k$ of DDIM in LCD is set to 10. For the CoMoSVC and DiffSVC models, it is necessary to train a separate vocoder that accepts mel-spectrogram as inputs. Therefore, we exploit the same structure as the BigVGAN in So-VITS-SVC to train a vocoder with NSF with the number of iterations set to 400k.
\subsection{Evaluation Metrics}
To construct the conversion trials, we separately establish pairs of conversions for both seen and unseen scenarios. A cross-validation strategy is employed to conduct audio clips where each singer provides the vocal content information as the source and other singers provide timbre information as targets. Consequently, 528 and 448 converted song clips are formed as the test trials for the seen and unseen cases respectively.

We evaluate these converted audios using objective metrics i.e. Singer Similarity (SSIM), F0 Pearson correlation (FPC), Mel Cepstral Distortion (MCD) and  Real-Time Factor (RTF). Specifically, we use the time ratio of the time taken to convert singing voices to the duration of audio as a representation of RTF.
All the inference processes are conducted on one NVIDIA A100 GPU. For subjective evaluation, we randomly pick 20 audio samples from each model in both seen and unseen situations. We employ a 5-point Mean Opinion Score (MOS) (1-bad, 2-poor, 3-fair, 4-good, 5-excellent) test and invite 20 volunteers to participate in the listening test, where both  Naturalness MOS (NMOS) and Similarity MOS (SMOS) are evaluated.
\begin{table}[t]
\caption{Objective metrics of different methods. The suffix 'S' represents the Seen Singer scenario, while the suffix 'U' represents the Unseen Singer scenario.}
\setlength\tabcolsep{2pt} 
\centering
\resizebox{0.48\textwidth}{!}{%
\begin{tabular}{l|c|c|c|c|c}
\toprule
\multicolumn{1}{l}{Method} &
  \multicolumn{1}{|c}{SSIM-S} &
  \multicolumn{1}{|c}{SSIM-U} &
  \multicolumn{1}{|c}{MCD-S} &
  \multicolumn{1}{|c}{MCD-U} &
  \multicolumn{1}{|c}{RTF} 
  \\ \midrule
DiffSVC    & $0.593$ & $ 0.586 $ & $\bf{3.466}$ & $\bf{3.574}$ &$0.166$ \\
So-VITS-SVC~~~ & $0.639$ & $ 0.600 $ &$3.960$ & $4.066$&$0.007$ \\
CoMoSVC  & $0.639$ & $ 0.600 $ &$3.944$ & $3.822$ &$\bf{0.004}$ \\
 LCM-SVC-T& $\bf{0.702}$ &$ \bf{0.663} $ & $3.819$& $4.072$ &$0.369$ \\
 LCM-SVC-4 & $0.697$ &  $ 0.658 $ &$3.769$& $4.012$ &$0.010$ \\
LCM-SVC-2 & $0.696$ &  ~$0.657$~ & ~$3.866$~ & ~$4.096$~ &~$0.007$~ \\
LCM-SVC-1 &  $0.690 $&  $ 0.652 $ & $4.040$& $ 4.252 $ &$\bf{0.004}$ \\
 \bottomrule
\end{tabular}
}
\label{obj}
\end{table}
\subsection{Experimental Results}
Table 1 shows the different results for seen and unseen scenarios. LCM-SVC-T significantly outperforms other models in terms of both NMOS and SMOS. After LCD, the performance of LCM-SVC has a slight loss, but still surpasses other baseline models in terms of timbre similarity, and the NMOS maintains  a  comparable level to the other models.

Table 2 shows the performance of LCM-SVC under multi-step conditions. It can be observed that in the seen case, an increase in the number of iterations significantly improves the sound quality, but has little effect on timbre. Under the unseen condition, an increase in the number of iterations greatly enhances both sound quality and timbre. Overall, these indicate that under the seen condition, the use of LCD tends to result in a reduction in sound quality, while in the unseen case it leads to a decrease in the timbre similarity.

Table 3 shows the SSIM, MCD and RTF of comparison methods under the unseen condition. It can be seen that the SSIM of the LCM-SVC methods is significantly higher than that of other models, but the MCD is at a higher level. This may be because both DiffSVC and CoMoSVC predict the Mel-spectrogram as the target, while So-VITS-SVC and LCM-SVC predict the latent representation. However, they are expected to have a better listening experience.
As for the crucial RTF metric, the RTF of the diffusion-based methods DiffSVC and LCM-SVC-T is at a higher level, which is undoubtedly disastrous for real-time applications. After LCD, the RTF of LCM-SVC-1 drops to 0.004, requiring only one-step inference, but its sound quality is consequently compromised. LCM-SVC-2 and LCM-SVC-4, on the other hand, can achieve a quite good sound quality and timbre performance at an acceptable RTF. This means that unless extreme RTF is required, using 2-step or 4-step inference can be a better choice in practice.

\section{Conclusion}

In this paper, we proposed the LCM-SVC, an any-to-any singing voice conversion method using the latent consistency model. By using the LCD method to distill a pre-trained LDM, we can reduce inference steps to one or much less, significantly decreasing the inference time. Tests on the OpenSinger dataset show that 1-step inference with LCM-SVC results in a performance drop, while the 2-step or 4-step inference yields performance comparable to the original LDM. It would be better to use a multi-step inference in situations where extreme low latency isn't necessary to maintain a high performance in real-time SVC applications.


\newpage
\bibliographystyle{IEEEtran}

\bibliography{mybib}

\begin{thebibliography}{10}
\providecommand{\url}[1]{#1}
\csname url@samestyle\endcsname
\providecommand{\newblock}{\relax}
\providecommand{\bibinfo}[2]{#2}
\providecommand{\BIBentrySTDinterwordspacing}{\spaceskip=0pt\relax}
\providecommand{\BIBentryALTinterwordstretchfactor}{4}
\providecommand{\BIBentryALTinterwordspacing}{\spaceskip=\fontdimen2\font plus
\BIBentryALTinterwordstretchfactor\fontdimen3\font minus \fontdimen4\font\relax}
\providecommand{\BIBforeignlanguage}[2]{{%
\expandafter\ifx\csname l@#1\endcsname\relax
\typeout{** WARNING: IEEEtran.bst: No hyphenation pattern has been}%
\typeout{** loaded for the language `#1'. Using the pattern for}%
\typeout{** the default language instead.}%
\else
\language=\csname l@#1\endcsname
\fi
#2}}
\providecommand{\BIBdecl}{\relax}
\BIBdecl

\bibitem{bytesing}
Y.~Gu, X.~Yin, Y.~Rao, Y.~Wan, B.~Tang, Y.~Zhang, J.~Chen, Y.~Wang, and Z.~Ma, ``{ByteSing}: A {Chinese} singing voice synthesis system using duration allocated encoder-decoder acoustic models and wavernn vocoders,'' in \emph{2021 12th International Symposium on Chinese Spoken Language Processing (ISCSLP)}.\hskip 1em plus 0.5em minus 0.4em\relax IEEE, 2021, pp. 1--5.

\bibitem{cui2024sifisinger}
J.~Cui, Y.~Gu, C.~Weng, J.~Zhang, L.~Chen, and L.~Dai, ``Sifisinger: A high-fidelity end-to-end singing voice synthesizer based on source-filter model,'' in \emph{ICASSP 2024-2024 IEEE International Conference on Acoustics, Speech and Signal Processing (ICASSP)}.\hskip 1em plus 0.5em minus 0.4em\relax IEEE, 2024, pp. 11\,126--11\,130.

\bibitem{polyak2020unsupervised}
A.~Polyak, L.~Wolf, Y.~Adi, and Y.~Taigman, ``Unsupervised cross-domain singing voice conversion,'' \emph{arXiv preprint arXiv:2008.02830}, 2020.

\bibitem{liu2021fastsvc}
S.~Liu, Y.~Cao, N.~Hu, D.~Su, and H.~Meng, ``Fastsvc: Fast cross-domain singing voice conversion with feature-wise linear modulation,'' in \emph{2021 ieee international conference on multimedia and expo (icme)}.\hskip 1em plus 0.5em minus 0.4em\relax IEEE, 2021, pp. 1--6.

\bibitem{zhou2022hifi}
Y.~Zhou and X.~Lu, ``{HiFi-SVC}: Fast high fidelity cross-domain singing voice conversion,'' in \emph{ICASSP 2022-2022 IEEE International Conference on Acoustics, Speech and Signal Processing (ICASSP)}.\hskip 1em plus 0.5em minus 0.4em\relax IEEE, 2022, pp. 6667--6671.

\bibitem{luo2020singing}
Y.-J. Luo, C.-C. Hsu, K.~Agres, and D.~Herremans, ``Singing voice conversion with disentangled representations of singer and vocal technique using variational autoencoders,'' in \emph{ICASSP 2020-2020 IEEE International Conference on Acoustics, Speech and Signal Processing (ICASSP)}.\hskip 1em plus 0.5em minus 0.4em\relax IEEE, 2020, pp. 3277--3281.

\bibitem{liu2021diffsvc}
S.~Liu, Y.~Cao, D.~Su, and H.~Meng, ``Diff{SVC}: A diffusion probabilistic model for singing voice conversion,'' in \emph{2021 IEEE Automatic Speech Recognition and Understanding Workshop (ASRU)}.\hskip 1em plus 0.5em minus 0.4em\relax IEEE, 2021, pp. 741--748.

\bibitem{lu2024comosvc}
Y.~Lu, Z.~Ye, W.~Xue, X.~Tan, Q.~Liu, and Y.~Guo, ``Comosvc: Consistency model-based singing voice conversion,'' \emph{arXiv preprint arXiv:2401.01792}, 2024.

\bibitem{chen2024ldmsvc}
S.~Chen, Y.~Gu, J.~Zhang, N.~Li, R.~Chen, L.~Chen, and L.~Dai, ``{LDM-SVC: Latent Diffusion Model Based Zero-Shot Any-to-Any Singing Voice Conversion with Singer Guidance},'' \emph{arXiv preprint arXiv:2406.05325}, 2024.

\bibitem{song2020denoising}
J.~Song, C.~Meng, and S.~Ermon, ``Denoising diffusion implicit models,'' \emph{arXiv preprint arXiv:2010.02502}, 2020.

\bibitem{lu2022dpm}
C.~Lu, Y.~Zhou, F.~Bao, J.~Chen, C.~Li, and J.~Zhu, ``Dpm-solver: A fast ode solver for diffusion probabilistic model sampling in around 10 steps,'' \emph{Advances in Neural Information Processing Systems}, vol.~35, pp. 5775--5787, 2022.

\bibitem{song2023consistency}
Y.~Song, P.~Dhariwal, M.~Chen, and I.~Sutskever, ``Consistency models,'' in \emph{Proceedings of the 40th International Conference on Machine Learning}, 2023, pp. 32\,211--32\,252.

\bibitem{luo2023latent}
S.~Luo, Y.~Tan, L.~Huang, J.~Li, and H.~Zhao, ``Latent consistency models: Synthesizing high-resolution images with few-step inference,'' \emph{arXiv preprint arXiv:2310.04378}, 2023.

\bibitem{kim2021conditional}
J.~Kim, J.~Kong, and J.~Son, ``Conditional variational autoencoder with adversarial learning for end-to-end text-to-speech,'' in \emph{International Conference on Machine Learning}.\hskip 1em plus 0.5em minus 0.4em\relax PMLR, 2021, pp. 5530--5540.

\bibitem{oord2016wavenet}
A.~v.~d. Oord, S.~Dieleman, H.~Zen, K.~Simonyan, O.~Vinyals, A.~Graves, N.~Kalchbrenner, A.~Senior, and K.~Kavukcuoglu, ``Wavenet: A generative model for raw audio,'' \emph{arXiv preprint arXiv:1609.03499}, 2016.

\bibitem{vaswani2017attention}
A.~Vaswani, N.~Shazeer, N.~Parmar, J.~Uszkoreit, L.~Jones, A.~N. Gomez, {\L}.~Kaiser, and I.~Polosukhin, ``Attention is all you need,'' \emph{Advances in neural information processing systems}, vol.~30, 2017.

\bibitem{papamakarios2021normalizing}
G.~Papamakarios, E.~Nalisnick, D.~J. Rezende, S.~Mohamed, and B.~Lakshminarayanan, ``Normalizing flows for probabilistic modeling and inference,'' \emph{Journal of Machine Learning Research}, vol.~22, no.~57, pp. 1--64, 2021.

\bibitem{lee2022bigvgan}
S.-g. Lee, W.~Ping, B.~Ginsburg, B.~Catanzaro, and S.~Yoon, ``Bigvgan: A universal neural vocoder with large-scale training,'' \emph{arXiv preprint arXiv:2206.04658}, 2022.

\bibitem{wang2019neural}
X.~Wang, S.~Takaki, and J.~Yamagishi, ``Neural source-filter waveform models for statistical parametric speech synthesis,'' \emph{IEEE/ACM Transactions on Audio, Speech, and Language Processing}, vol.~28, pp. 402--415, 2019.

\bibitem{ho2020denoising}
J.~Ho, A.~Jain, and P.~Abbeel, ``Denoising diffusion probabilistic models,'' \emph{Advances in neural information processing systems}, vol.~33, pp. 6840--6851, 2020.

\bibitem{wu2021cross}
P.~Wu, J.~Pan, C.~Xu, J.~Zhang, L.~Wu, X.~Yin, and Z.~Ma, ``Cross-speaker emotion transfer based on speaker condition layer normalization and semi-supervised training in text-to-speech,'' \emph{arXiv preprint arXiv:2110.04153}, 2021.

\bibitem{ho2022classifier}
J.~Ho and T.~Salimans, ``Classifier-free diffusion guidance,'' \emph{arXiv preprint arXiv:2207.12598}, 2022.

\bibitem{huang2021multi}
R.~Huang, F.~Chen, Y.~Ren, J.~Liu, C.~Cui, and Z.~Zhao, ``Multi-singer: Fast multi-singer singing voice vocoder with a large-scale corpus,'' in \emph{Proceedings of the 29th ACM International Conference on Multimedia}, 2021, pp. 3945--3954.

\bibitem{radford2023robust}
A.~Radford, J.~W. Kim, T.~Xu, G.~Brockman, C.~McLeavey, and I.~Sutskever, ``Robust speech recognition via large-scale weak supervision,'' in \emph{International Conference on Machine Learning}.\hskip 1em plus 0.5em minus 0.4em\relax PMLR, 2023, pp. 28\,492--28\,518.

\bibitem{kim2018crepe}
J.~W. Kim, J.~Salamon, P.~Li, and J.~P. Bello, ``Crepe: A convolutional representation for pitch estimation,'' in \emph{2018 IEEE International Conference on Acoustics, Speech and Signal Processing (ICASSP)}.\hskip 1em plus 0.5em minus 0.4em\relax IEEE, 2018, pp. 161--165.

\bibitem{wang2023cam++}
H.~Wang, S.~Zheng, Y.~Chen, L.~Cheng, and Q.~Chen, ``{CAM++}: A fast and efficient network for speaker verification using context-aware masking,'' \emph{arXiv preprint arXiv:2303.00332}, 2023.

\end{thebibliography}

\end{document}